\PassOptionsToPackage{dvipsnames,svgnames}{xcolor}

\documentclass[conference, 10pt]{IEEEtran}

\usepackage{subcaption}
\usepackage{tabulary}
\usepackage{threeparttable}
\usepackage{booktabs}
\usepackage{relsize,xspace}
\usepackage{multirow}
\usepackage{enumitem}
\usepackage{balance}
\usepackage{array}
\usepackage{hyperref}
\usepackage{multicol}
\usepackage{microtype}
\usepackage[table,xcdraw]{xcolor}
\usepackage{makecell}
\usepackage{graphicx}
\usepackage[capitalize]{cleveref}
\usepackage[textsize=tiny]{todonotes}
\usepackage{amssymb}
\usepackage{tabularx,cellspace}
\usepackage{makecell}
\usepackage{adjustbox}

\newcommand\parhead[1]{\vspace{.5mm}\noindent\textbf{{#1}}}

\newcommand{\quotebox}[3]{\vspace{.5em}\noindent\begin{tikzpicture}
    \node[align=center,draw,thin,minimum width=\columnwidth,inner sep=1.5mm] (titlebox)%
    {\parbox{\dimexpr \linewidth-2\fboxsep-2\fboxrule}{\looseness=-1\noindent\textit{#2\vspace{0.1cm}}}};
    \node[label=left:{\colorbox{white}{\small #1}}] (W) at (titlebox.south east) {};%
    \end{tikzpicture}\vspace{-15pt}}

\newcommand{\sumbox}[1]{
        \noindent\fbox{\parbox{\dimexpr \linewidth-2\fboxsep-2\fboxrule}{\noindent
                \textit{#1}
            }}\smallskip
        }

\begin{document}
\newcommand{\featurename}{SSF} 
\newcounter{feature}
\crefformat{feature}{#2\featurename\textsubscript{#1}#3}
\Crefformat{feature}{#2\featurename\textsubscript{#1}#3}
\newcommand{\feature}[1]{\refstepcounter{feature}\textbf{\featurename\textsubscript{\thefeature}: }\label{#1}}

\newcommand{\characteristicname}{CSF} 
\newcounter{characteristic}
\crefformat{characteristic}{#2\characteristicname\textsubscript{#1}#3}
\Crefformat{characteristic}{#2\characteristicname\textsubscript{#1}#3}
\newcommand{\characteristic}[1]{\refstepcounter{characteristic}\textbf{\characteristicname\textsubscript{\thecharacteristic}: }\label{#1}}

\newcommand{\engineeringname}{EPSF} 
\newcounter{engineering}
\crefformat{engineering}{#2\engineeringname\textsubscript{#1}#3}
\Crefformat{engineering}{#2\engineeringname\textsubscript{#1}#3}
\newcommand{\engineering}[1]{\refstepcounter{engineering}\textbf{\engineeringname\textsubscript{\theengineering}: }\label{#1}}

\newcommand{\challengename}{CH} 
\newcounter{challenge}
\crefformat{challenge}{#2\challengename\textsubscript{#1}#3}
\Crefformat{challenge}{#2\challengename\textsubscript{#1}#3}
\newcommand{\challenge}[1]{\refstepcounter{challenge}\textbf{\challengename\textsubscript{\thechallenge}: }\label{#1}}

\newcommand{\cut}[1]{}
\newcommand\edit[1]{{\color{black}#1}}

\newcommand{\tb}[1]{{\textsf{\textbf{TB}}[\smaller\sffamily\color{orange} #1]}}
\newcommand{\svp}[1]{{\textsf{\textbf{SP}}[\smaller\sffamily\color{yellow} #1]}}
\newcommand{\kh}[1]{{\textsf{\textbf{KH}}[\smaller\sffamily\color{blue} #1]}}
\newcommand{\rwi}[1]{{\textsf{\textbf{RW1}}[\smaller\sffamily\color{red} #1]}}
\newcommand{\rwii}[1]{{\textsf{\textbf{RW2}}[\smaller\sffamily\color{red} #1]}}
\newcommand{\rwiii}[1]{{\textsf{\textbf{RW3}}[\smaller\sffamily\color{red} #1]}}

\definecolor{darkorange}{rgb}{0.8, 0.55, 0.0}
\newcommand*{\result}[3]{\noindent\textbf{\mbox{\textcolor{DarkGreen}{#1~agree};} \mbox{\textcolor{DarkRed}{#2~disagree}; \mbox{\textcolor{darkorange}{#3~not mentioned}}}}\vspace{1.5pt}}
\newcommand*{\occurences}[1]{\noindent\textbf{\mbox{\textcolor{DarkGreen}{(#1~occurences)}}}\vspace{1.5pt}}



\title{An Exploratory Study on\\the Engineering of Security Features}









 \author{
 \IEEEauthorblockN{
     Kevin Hermann\IEEEauthorrefmark{1}, Sven Peldszus\IEEEauthorrefmark{1}, Jan-Philipp Steghöfer\IEEEauthorrefmark{2}, Thorsten Berger\IEEEauthorrefmark{1}\IEEEauthorrefmark{3}
 }
  	\IEEEauthorblockA{
         \IEEEauthorrefmark{1}Ruhr University Bochum, Germany \hspace{.3cm}%
         \IEEEauthorrefmark{2}XITASO GmbH, Germany\hspace{.3cm}%
         \IEEEauthorrefmark{3}Chalmers\,$|$\,University of Gothenburg, Sweden 
  	}
      }

\maketitle
\begin{abstract}
	Software security is of utmost importance for most software systems. Developers must systematically select, plan, design, implement, and especially, maintain and evolve security features---functionalities to mitigate attacks or protect personal data such as cryptography or access control---to ensure the security of their software. Although security features are usually available in libraries, integrating security features requires writing and maintaining additional security-critical code. While there have been studies on the use of such libraries, surprisingly little is known about how developers engineer security features, how they select what security features to implement and which ones may require custom implementation, and the implications for maintenance. As a result, we currently rely on assumptions that are largely based on common sense or individual examples. However, to provide them with effective solutions, researchers need hard empirical data to understand what practitioners need and how they view security---data that we currently lack. To fill this gap, we contribute an exploratory study with 26 knowledgeable industrial participants. We study how security features of software systems are selected and engineered in practice, what their code-level characteristics are, and what challenges practitioners face. Based on the empirical data gathered, we provide insights into engineering practices and validate four common assumptions.


\end{abstract}

\begin{IEEEkeywords}
    Security Feature, Software Security, Secure Software Development, Security by Design, Developer Study
\end{IEEEkeywords}

\section{Introduction}
\noindent\looseness=-1
Considering security in every development phase is a critical\,\cite{McGraw2004} yet challenging task.
To secure a software system, developers must engineer its security features.
Security features, such as encryption or access control, address security concerns by mitigating an attack or protecting assets, such as personal data, to prevent malicious actions of an attacker\,\cite{McGraw2004}.
Security features must be selected according to the system's security objectives, for example, according to the CIA triad\,\cite{Cawthra2020} or the Parkerian Hexad\,\cite{Parker2015}.
Following the literature, the engineering of security features should start with the identification of security requirements\,\cite{Sindre2005,Haley2008,Mellado2010}, continue with  secure design\,\cite{Juerjens05,shostack2014threat,SDG2017,casola2020securityautomation} and implementation\,\cite{green2016developers}, and end with validation and verification of the realized features\,\cite{potter2004,PTS+2019,TPS2022,di2023}.

\looseness=-1
The research community invested substantial effort in developing security technologies and engineering techniques to keep pace with adversarial actors who continuously discover new vulnerabilities.
Most developers, however, are not security experts\,\cite{Ryan2023,Wurster2008,green2016developers}.
While libraries are available for a wide range of security features, developers still have to write a notable amount of security-critical code to configure and integrate them\,\cite{KNR+2017,WBBM2021,egele2013empirical}, which often leads to insecure implementation of security features despite the use of security libraries\,\cite{Acar2017}.
Further difficulties arise when developers must implement custom security features for which no library is available.

\looseness=-1
Research so far has focused on individual aspects of the development of security features in isolation\,\cite{egele2013empirical,Oyetoyan2016,nadi2016jumping,green2016developers,Acar2017,Venson2019,WBBM2021}, such as investigating the usability of cryptographic APIs and its impact on security in controlled experiments\,\cite{Acar2017}.
\textit{However, we still lack a holistic understanding of how developers engineer security features in practice and what the \edit{code-level characteristics} of security features are.}
The absence of hard empirical data has lead to
\textit{individual, anecdotal views on the engineering of security features in software projects}, and
\textit{research commonly makes assumptions that might not always be accurate}, as a recent study shows\,\cite{Ryan2023}.

\looseness=-1
To improve software security and base research on valid assumptions, we need a better understanding of how practitioners engineer security features.
\edit{In particular, we need to know on which basis security features are selected and why some are not implemented.
Furthermore, we require an understanding of how security features are engineered in practice, the challenges involved, as well as the characteristics of security features.}
We identified the following research questions to address this gap:
\smallskip
\begin{description}
    \item [\textbf{RQ\textsubscript{1}:}] {What is the developer perspective on security features, and what influences security feature selection?}%
    \smallskip
    \item [\textbf{RQ\textsubscript{2}:}] {How is the engineering of security features embedded into the software development lifecycle?}%
    \smallskip
    \item [\textbf{RQ\textsubscript{3}:}] {What are code-level characteristics of security features?}%
    \smallskip
    \item [\textbf{RQ\textsubscript{4}:}] {What challenges arise in engineering security features?}
\end{description}
\smallskip
\looseness=-1
We present an exploratory study using semi-struc\-tured interviews with 26 industry experts.
We provide the interview guide, aggregated data, and the evaluation scripts to derive conclusions in our replication packge\,\cite{ReplicationPackage}.
The interviews provide insights into practices for engineering security features in software systems.
We reason about four common assumptions in security research, combining insights from our interviews and observations from existing studies.

\looseness=-1
We confirm several commonly held assumptions while revealing significant room for improvement.
In particular, we confirm that developers lack \edit{knowledge related to detailed security concepts, but are still able to realize security features on a practical level.} 
As a result, many developers struggle with secure configuration of security libraries. 
Contrary to popular belief, security-by-design techniques are used in practice, but mainly in regulated domains.
In general, however, most companies try to follow security-by-design principles, but are forced to prioritize system functionality over security features.


\section{Background and Related Work}
\label{sec:background}
\noindent
\edit{Outside of the security domain, several definitions of ``feature'' focus on the} distinction between functionalities\,\cite{berger2015feature,bosch2000} and products\,\cite{Batory:2004bw}, or customer visibility\,\cite{chen2005,kang1990} and value\,\cite{riebisch2003}.
Related to this, \emph{Security features} provide functionalities that address security issues by preventing attacks on a software system\,\cite{McGraw2004}.
They either realize non-functional security requirements\,\cite{potter2004} \edit{or functional ones\,\cite{hermann2025}, e.g., to handle authentication, access control, cryptography}, and other aspects of software security\,\cite{tsipenyuk2005}.
\edit{To this end, security features aim to resist, detect or recover from attacks to a software system\,\cite{Bass2003}, or increase its resistance, tolerance, or resilience against them\,\cite{Allen2008}.}

\edit{To realize security features, Schumacher et al.\,\cite{schumacher2013} list security patterns for different stages of the software development lifecycle such as validation, threat assessment, or risk assessment.
An interview study on security testing revealed that practitioners recommend forming dedicated and specialized teams to handle security testing\,\cite{di2023}.
Oyetoyan et al.\,\cite{Oyetoyan2016} compared the security engineering of two agile organizations, finding that knowledge transfer is essential to increase security but must be actively approached.}
One of the main challenges is the significant difference in granularity at which security features can be considered\,\cite{Peldszus2022}%
\edit{, i.e., abstract planning of which information needs to be protected by which type of security feature, as opposed to detailed security feature characteristics such as their configuration.}
While previous work investigated the extent of implemented security features in open source projects through a keyword search\,\cite{ryoo2016}, no study investigated the \edit{code-level} characteristics of security features, such as scattering\,\cite{passos2015scattering} or tangling\,\cite{apel2013featureinteractions}.

Ryan et al.~\cite{Ryan2023} performed a systematic mapping study to identify common assumptions in security research containing contradictory findings.
Unlike our study, they do not generate new empirical data, but systematically relate existing data to identify misconceptions, serving as a primary motivation for our study.
\edit{In 2009, Werlinger et al.\,\cite{Werlinger2009} identified human, organizational, and technological challenges of IT security management related to security experience, prioritization, and tools.
Our study reports on which methods practitioners employ to overcome such challenges, and which challenges remain.
Klivan et al.\,\cite{Klivan2024} found that developers of open source projects rarely communicate their security practices to other contributors, but expect them to utilize sensible ones.
However, it is unclear if this also applies to companies as well.}

\section{Methodology}
\label{sec:methodology}
\noindent
%
\edit{
	\noindent
	We first selected assumptions security researchers make about security engineering in practice, followed by interviews with practitioners to gain insights on the engineering process.
}

\subsection{Selection of Assumptions}
\noindent
\edit{
	To identify common assumptions in security research, we contacted 39 security researchers from 14 institutions and asked them about \emph{``assumptions about security in practice they typically encounter in their research.''}
	We received responses both in discussion and in writing from 22 researchers from nine institutions in various fields. 
	We grouped similar responses and triangulated them with related work to derive and detail the assumptions about security in~practice.
	We then formulated research questions to validate these assumptions.
}

\subsection{Interview Study}
\noindent\looseness=-1
We conducted semi-structured interviews\,\cite{shull2007,denzin2011} with domain experts to obtain insights about engineering security features.

\parhead{Recruitment.}
We recruited developers, software architects, security engineers, and project managers involved in at least one phase of the engineering process of security features.
We recruited participants through our personal networks, and by asking computer science associations and participants for referrals until reaching saturation\,\cite{Weller2018}.
We treated saturation as being able to understand the meaning of each code in the coding process\,\cite{Hennink2017}.
After interviewing 17 participants, we reached saturation, which is consistent with the experience of Hennink et al.\,\cite{Hennink2017} of 16 to 24 interviews.
To mitigate the risk of saturation by chance, we conducted further interviews.

\begin{table}
	\centering
	\caption{\edit{Interview participants and their characteristics}}
	\label{tab:participants}
	\setlength{\tabcolsep}{3.2pt}
	\smaller
\begin{threeparttable}
	\begin{tabular}{rlrl}
		\toprule
		\textbf{ID}\hspace{3pt} & \textbf{Role} & \textbf{Experience} & \textbf{Domain} \\
		\midrule
		I1\hspace{3pt} & Software Architect & 5 years & Logistics \\
		I2\hspace{3pt} & Project Manager & 8 years & Quantum Computing \\
		I3\hspace{3pt} & Software Developer & 5 years & Antivirus \\
		I4\hspace{3pt} & Project Manager & 22 years & Automotive Security \\
		I5\hspace{3pt} & Security Engineer & 4 years & Insurance, Public Sector \\
		I6\hspace{3pt} & Software Developer & 7 years & Real Estate \\
		I7\hspace{3pt} & Security Engineer & 8 years & Insurance, Banking, Retail \\
		I8\hspace{3pt} & Software Developer & 7 years & Logistics \\
		I9\hspace{3pt} & Software Developer & 10 years & Insurance \\
		I10\hspace{3pt} & Software Architect & 6 years & Medical, Automotive, Robotics \\
		I11\hspace{3pt} & Software Developer & 5 years & Systems Management \\
		I12\footnotemark[1] & Security Engineer & 6 years & Automotive Security \\
		I13\footnotemark[1] & Security Engineer & 4 years & Automotive Security \\
		I14\footnotemark[1] & Security Engineer & 6 years & Automotive Security \\
		I15\hspace{3pt} & Software Developer & 11 years & Medical, Automotive \\
		I16\hspace{3pt} & Software Developer & 8 years & Manufacturing \\
		I17\hspace{3pt} & Software Developer & 4 years & Medical, Insurance \\
		I18\hspace{3pt} & Software Developer & 5 years & Logistics \\
		I19\hspace{3pt} & Software Architect & 9 years & Banking \\
		I20\hspace{3pt} & Software Developer & 15 years & Cloud Computing \\
		I21\hspace{3pt} & Project Manager & 20 years & Cloud Computing \\
		I22\hspace{3pt} & Software Architect & 8 years & Insurance \\
		I23\hspace{3pt} & Software Developer & 10 years & Machine Learning, Cloud Computing \\
		I24\hspace{3pt} & Software Architect & 15 years & Infrastructure, Aerospace, Defense \\
		I25\hspace{3pt} & Software Developer & 7 years & Manufacturing \\
		I26\hspace{3pt} & Software Developer & 3 years & Manufacturing \\
		\bottomrule
	\end{tabular}
	\begin{tablenotes}
	\item \footnotemark[1]I12--I14 were part of one focus group
	\end{tablenotes}
\end{threeparttable}
\vspace{-19pt}
\end{table}

\begin{figure}
	\centering
	\begin{subfigure}{.9\linewidth}
		\includegraphics[width=\linewidth]{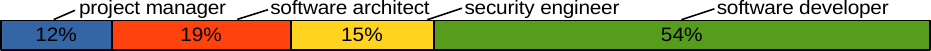}
		\caption{Roles of the participants}
	\end{subfigure}
	\smallskip
	\begin{subfigure}{.9\linewidth}
		\includegraphics[width=\linewidth]{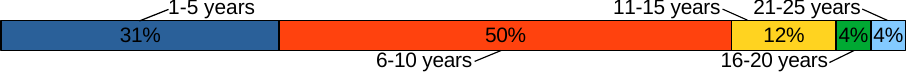}
		\caption{Experience of the participants in their current role}
	\end{subfigure}
	\smallskip
	\begin{subfigure}{.9\linewidth}
		\includegraphics[width=\linewidth]{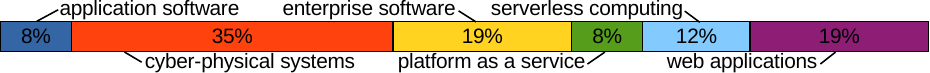}
		\caption{Kinds of developed software systems}
	\end{subfigure}
	\smallskip
	\begin{subfigure}{.9\linewidth}
		\includegraphics[width=\linewidth]{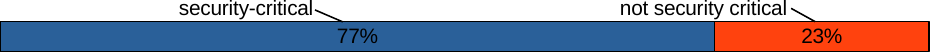}
		\caption{Criticality of the developed systems}
	\end{subfigure}
	\vspace{-6pt}
	\caption{Overview of participants for the interviews}
	\label{fig:participants}
	\vspace{-6pt}
\end{figure}

\looseness=-1
\parhead{Participants.}
\Cref{tab:participants,fig:participants} summarize the participants.
We interviewed 26 professionals from 15 companies covering projects across various domains. Seven participants work as consultants for external companies.
Six participants have experience from multiple domains.
Some companies work on a single product, others build customized software.
All interviewed project managers have a strong background in software engineering, while one has a security background.
On average, participants have 8.4 years of experience in their role, ranging from 3 to 22 years.
Participants work on diverse types of software systems, with cyber-physical systems, enterprise software, and web applications being the most prominent.
We conducted one interview as a focus group with three participants.

\begin{figure}
	\centering
	\includegraphics[width=\linewidth]{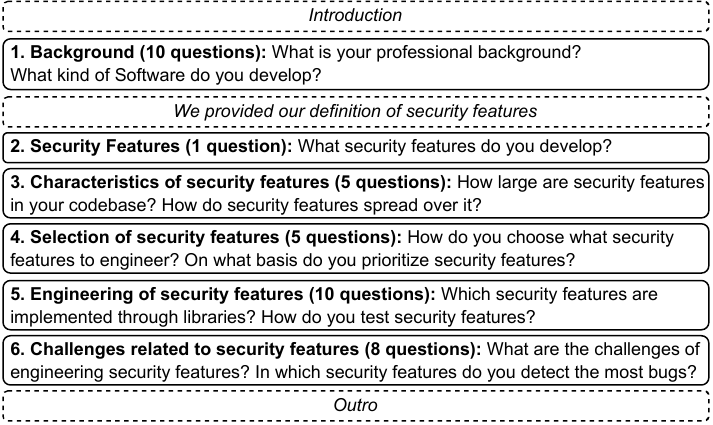}
	\caption{\edit{Outline of the interview guide with sample questions.}}
	\label{fig:interview_outline}
	\vspace{-15pt}
\end{figure}

\parhead{Interview Design.}
Four researchers iteratively drafted the interview questions by proposing various questions according to the research questions.
The questions were then thoroughly discussed among the researchers, merged, and refined multiple times to form the interview guide.
\Cref{fig:interview_outline} shows the resulting outline of our interview guide with sample questions.

\looseness=-1
\textbf{(1)} We began by asking the interviewees about their professional background, the software they develop, and the importance of security in their projects.
\textbf{(2)} We then explained, that ``a security feature is a feature implementing a specific kind of security mechanism to mitigate an attack or protect an asset such as personal data.'', and asked them to list security features important to them, \textbf{(3)} as well as their \edit{code-level} characteristics.
Then, we inquired about how they handle and communicate security features at each development stage, focusing on \textbf{(4)} selection, prioritization, \textbf{(5)} design, implementation, testing, and maintenance.
\textbf{(6)} Finally, we discussed challenges in engineering security features, and ways to overcome them.


\begin{figure}[b]
	\vspace{-20pt}
	\centering
	\includegraphics[width=\linewidth]{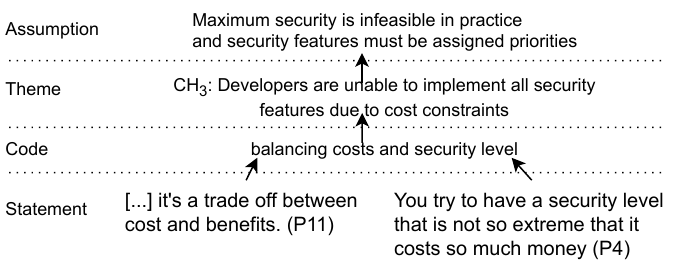}
	\caption{Illustration of the coding process}
	\label{fig:methodology}
\end{figure}

\parhead{Analysis.}
We recorded the interviews with consent of the participants.
The first author transcribed the recordings using Whisper\,\cite{Whisper}, run locally, and manually corrected inconsistencies between transcripts and recordings.
Interviews ranged in length from 34 to 95 minutes.
We stored all recordings and transcripts only on computers within our institution, only the authors had access to them, and we deleted them after analysis.

We applied open coding\,\cite{Boyatzis1998} to extract themes from the transcribed interviews using MAXQDA\,\cite{MAXQDA}.
\Cref{fig:methodology} illustrates the coding process.
The first author coded the transcripts to extract key information from the answers to the interview questions.
We then merged codes to a theme, which we mapped to the assumptions.
To ensure consistency, the first and second author discussed the code book in detail.

\looseness=-1
For each theme, we indicate the number of interviews in which it was substantiated or contradicted and in which it did not come up (\result{X}{Y}{Z}).
We also investigated the influence of roles or domains on each derived theme. 
We discuss these explicitly where observed.

\looseness=-1
\parhead{Replication Package.}
We provide our interview guide as well as our aggregated results in our replication package\,\cite{ReplicationPackage}.


\section{Common Assumptions in Security Research}
\label{sec:prestudy}
\noindent
\edit{We identified 4 common assumptions that researchers frequently make and count how often they occurred in their responses.

\parhead{Assumption 1: Ordinary developers lack knowledge to securely engineer security features} \occurences{10}

\noindent\looseness=-1
\edit{
	Many academics make assumptions about developers' security knowledge and ability to perform certain tasks, e.g., security planning or risk assessment.
	On the one hand, developers are rarely considered to be security experts and, therefore, need tools to help them plan and implement security features\,\cite{Wurster2008,green2016developers}.
	On the other hand, developers are assumed to be capable of performing complex security tasks related to security feature engineering\,\cite{Ahmadian2017,Reiche2024}.
	Such conflicting views on the assumption that \textit{``all developers know how to code securely''} is also reflected in previous work\,\cite{Ryan2023}.
	However, we still lack a detailed understanding of developers' actual security knowledge.
}

\parhead{Assumption 2: (Model-based) security-by-design techniques are not used in practice} \occurences{7}

\noindent\looseness=-1
Various model-based security-by-design techniques have been proposed\,\cite{Krausz2024}, e.g., threat modeling based on data flow diagrams\,\cite{shostack2014threat,tuma2018threat,TSB2019} or various extensions to design languages such as UMLsec\,\cite{Juerjens05}.
The literature frequently reports benefits of the practical application of model-based security-by-design techniques\,\cite{Apvrille2005,Best2007,Juerjens2008,Basin2011}.
Yet, researchers frequently acknowledge that the adoption of such techniques requires models that may not be used in practice\,\cite{Seifermann2019,TSB2019,Peldszus2021,Peldszus2024}.
A recent empirical study\,\cite{Krausz2024} shows that most design-time security languages extend design models, and therefore, assume their use.
On the other hand, Gorschek et al.\ show that design models are rarely used in practice\,\cite{Gorschek2014}.

\parhead{Assumption 3: Security libraries are complicated to use, resulting in many vulnerabilities} \occurences{5}

\noindent\looseness=-1
Developers have access to a large set of libraries that include security features.
However, complex cryptographic APIs are commonly assumed to be a main reason for insecure code.
Developers often perceive these libraries as too low-level and to require a lot of manual implementation to be useful\,\cite{nadi2016jumping}.
As a result, they use security libraries in insecure ways, leading to security issues\,\cite{Acar2017}.
To counteract this, tools such as CogniCrypt\,\cite{KNR+2017} analyze the use of cryptographic APIs to detect insecure usages.
While studies draw a concrete picture for cryptographic APIs, their insights are often generalized. However, we lack concrete insights on what generally challenges practitioners when using security features provided by libraries.

\parhead{Assumption 4: Maximum security is infeasible in practice and security features must be assigned priorities} \occurences{7}

\noindent\looseness=-1
It is often assumed that \edit{organizations will make security a top priority} and apply research results\,\cite{Ryan2023}.
In practice, however, resource constraints often force organizations to make difficult trade-offs between security and functionality.
Since complete security is neither feasible nor desirable due to these constraints, prioritization between security features seems unavoidable\,\cite{Mazurek2022}, but has not been assessed empirically.
}
\section{\edit{Selection of Security Features and Influencing Factors (RQ\textsubscript{1})}}
\label{sec:selection}
\noindent
\looseness=-1
First, we focus on what participants consider a security feature and how security features are selected in practice.
\smallskip

\noindent
\textbf{\feature{importance} Security \edit{features are} generally perceived as important, but \edit{less important than functionality}}

\noindent\result{22}{4}{0}

\sumbox{Participants view security features as important for their projects, but customers often favor functional features. This creates challenges in prioritizing security features \edit{over functional ones}, particularly with limited resources.}\smallskip

\noindent
Six interviewees report frequently discussing the need to assign a higher priority to security features to mitigate costs when vulnerabilities are exploited with customers.
Customers, however, rarely take these security risks seriously, neglecting the security of the system which can cause vulnerabilities.

Seven participants mention customers not seeing security features as providing business value, thus prioritizing functional features over them: \textit{``functionality, performance, security. That's the order.''} (I6, Developer, Real Estate).
Two interviewees explain that in experimental domains, e.g., quantum computing, providing functionality is the prerequisite to securing it.

\smallskip

\noindent
\textbf{\feature{feature} Only few security features are of immediate concern}

\noindent\result{16}{6}{4}

\sumbox{Participants mentioned important security features, including cryptography, authentication, and authorization, while features such as validation and logging are less prominent.}\smallskip

\begin{table}
	\centering
	\caption{Security features mentioned more than once}
	\vspace{-3pt}
	\label{tab:features}
	\setlength{\tabcolsep}{3pt}
	\scriptsize
	\begin{tabularx}{\columnwidth}{r ccc X}
		\toprule
		\textbf{Security Feature} & \multicolumn{3}{c}{\textbf{Times Named}} & \textbf{Relation to OWASP Top 10}\\
		& {total} & {in\,step\,3} & {later} &  \\
		\midrule
		Authentication		& 18 & \textbf{14} &	4 & A07:Authentication\\
		Cryptography		& 17 & \textbf{12} & 5 & A02:Cryptography\\
		Authorization		& 16 & \textbf{10} & 6 & A01:Broken\,Access\,Control\\
		Validation			& 13 & 5  & \textbf{8} & A03:Injection,\,A08:Data\,Integrity\\
		Logging				& 8  & 2  & \textbf{6} & A09:Security\,Logging\\
		Security\,Monitoring & 7  & \textbf{6}  &	1 & A09:Security\,Monitoring\\
		Session\,Management 	& 3  & 1  &	\textbf{2} & A01:Access\,Control,\,A07:Authentication\\
		\bottomrule
	\end{tabularx}
\vspace{-10pt}
\end{table}

\noindent
\cref{tab:features} lists the security features mentioned in at least one interview\edit{, while also relating them to the OWASP Top 10 vulnerabilities}.
We gathered the security features mentioned in step 3 of our interview guide in \cref{fig:interview_outline}.
We also considered security features mentioned at a later stage of the interview, separating them from the initial response to identify potential patterns.
This distinction provides an indication that developers are immediately concerned with the security features \textit{authentication}, \textit{authorization}, and \textit{cryptography}.

While participants state to \textit{``definitely have authentication/authorization services in all of our software''} (I7, Security Engineer, Insurance, Banking, Retail), they perceive cryptography as for domains that handle personal information, such as real estate or banking, but less so in other domains such as logistics: \textit{``Encryption actually is a very rarely used part in the applications I'm working with [\ldots]
For example, if an external HTTPS API was used, then that's encryption, but you actually never get in touch with that because you just import the API library and do the API request and what it does internally, \edit{you never care about as a developer}''}\,(I8, Developer, Logistics).

Data validation, \textit{``is always required at the system boundaries''}\,(I1, Architect, Logistics) when systems receive input, especially in security patterns such as distrustful decomposition or zero trust architecture.
Similarly, eight participants highlight the importance of logging, e.g., of login attempts, accesses, or transactions, for accountability, i.e., to detect and understand security incidents.
Developers also implement log filtering to ensure that sensitive data is not leaked through logs.

\looseness=-1
Security monitoring to detect attacks early---reducing cost for mitigation and recovery---and session management are less frequently mentioned.
Some participants considered architectural patterns or refer to development activities as security features, which are not features within the system itself, such as firewalls around the company's intranet or information flow control, to prevent secrets in publicly accessible variables or objects.

\looseness=-1
Six vulnerabilities of the OWASP Top 10 can be directly related to the mentioned security features, although \textit{Broken Access Control}, as well as \textit{Cryptography and Authentication Failures} are of more immediate concern.
Note that the remaining four vulnerabilities, such as Insecure Design, are not directly related to a security feature, but loosely related to multiple ones.
\smallskip

\noindent
\textbf{\feature{experience} Security feature selection is based on experience}

\result{16}{0}{10}

\sumbox{%
	The selection of security features varies, ranging from unstructured to well-structured processes for capturing threats. The selection of security features is always based on experience, either from past projects or observations of competitors.
}
\smallskip

\noindent
Participants report, feature selection \edit{\textit{``is based on years of working with customers and knowing what's needed''} (I21, Architect, Banking)}.
Four participants mention, when not structuring security engineering processes, companies select security features based on those from other systems.

When security features stem from dedicated processes, such as threat modeling\,(cf.\,\cref{threat-model}), concrete security features are still discussed and selected based on experience, as noted in two interviews.
Required structured risk assessment processes in regulated domains\,\cite{ISO21434} are mentioned by three participants.

Finally, seven participants state that development teams discuss which security features are needed in team discussions and workshops which include developers and security experts.
\smallskip

\noindent
\textbf{\feature{specialization} Developers \edit{perceive security as an extensive domain and therefore} specialize in certain security features}

\result{15}{2}{9}

\sumbox{%
	Developers specialize in specific security features, as knowing all details about every security feature is infeasible, and act as ``lighthouses'' to help others in understanding these features.
}\smallskip

\noindent
\looseness=-1
More than half of the interviewees claim that it is infeasible to learn about every single security feature, \edit{\textit{``because you can't be an expert in all things''} (I26, Developer, Manufacturing)}. Instead, they focus on specific security features for which they become experts.
To reduce overhead, companies sometimes offer training or workshops on various security topics.
Typically, developers are intrinsically motivated to learn about security, becoming experts (often called ``lighthouses'') who assist other developers when working with specific security aspects.
\smallskip

\noindent
\textbf{\feature{prioritize} Customers are a main driver when companies need to assign priorities to security features}

\result{18}{3}{5}

\sumbox{%
	Except for absolutely necessary security features that must always be implemented, customers are a primary factor in decisions on extending them.
	Explicitly requested features that benefit most customers receive the highest priority.
}\smallskip

\noindent
\looseness=-1
Once a baseline of security features is implemented\,(cf.\,\cref{baseline}), the priority shifts to what benefits most customers or is actively requested, as revealed in fourteen interviews. 
Two participants mentioned that internal discussions and the impact on multiple customers help assign priorities to these features.


One interviewee explains that once security features have been selected, they are often realized in a logical order based on their interdependencies:
\textit{``if you don't have authentication, authorization doesn't make sense. You can enable authorization, but disable authentication. So I can just tell the system, 'hi, my name is [name]' and it trusts me. But obviously, in that case, I can circumvent authorization checks.''} (I21, \edit{Manager,~Cloud}).
\smallskip





\quotebox{Selection of Security Features (RQ1)}{Security feature selection is driven by experience, customer demand, and requirements from regulations. Few organizations follow structured processes to select security features, relying on customers to provide requirements and prioritization. However, they rarely provide these and favor the implementation of functional features. The most pressing security features are authentication and authorization. Validation and logging is common, in particular for architectural planning and accountability. To handle all these security features, developers become experts in a small subset of them.}{1.8cm}

\section{Engineering Security Features (RQ\textsubscript{2})}
\label{sec:engineering}
\noindent
In RQ\textsubscript{1}, we covered factors that influence the selection of security features.
Now, we focus on how security features embedded into software development processes.
\smallskip





\noindent
\edit{\textbf{\engineering{requirements} Projects usually lack initial security requirements}}

\noindent\result{19}{0}{7}

\sumbox{%
	Projects in unregulated domains rely on customers to provide security requirements or attack scenarios. They are rarely able to provide these, however, impeding security feature selection.
	Laws and regulations in certain domains are an important source of security requirements at project start.
}

\noindent
Many participants reported that customers, as non-experts, do not request specific security features.
Instead,
\textit{``They just have the idea that they have to protect their asset''} (I4, Manager, Automotive), as described in five interviews. This makes planning difficult due to often unclear concrete attack scenarios.
For example, one interviewee (I16, Developer, Manufacturing) mentions a customer who \textit{``wanted security''} for a medical device \textit{``but were unclear of which actual attack scenarios''} exist, as the device neither communicates nor stores data over a network and is only accessible by a doctor.
In such a case, determining attack scenarios is challenging as neither data nor the system itself is at risks.

\looseness=-1
Only participants working in domains regulated by laws and standards, such as automotive or banking, report starting projects with a well-defined list of required security features.
There, initial requirements of security features often stem from relevant laws and standards.
\edit{Since these requirements must be fulfilled for compliance, developers must assign a higher priority to these security features than in unregulated domains.
This reasoning was also provided by participants working in several domains with different levels of security criticality.}

\looseness=-1
\smallskip

\noindent
\textbf{\engineering{by-design} Companies try to follow security-by-design, but only rarely implement specific processes}

\result{25}{1}{0}

\sumbox{%
	Companies consider security as early as possible, but most do not follow a security-by-design process.
	Still, processes are needed to address previously neglected security concerns.
}\smallskip

\noindent
\looseness=-1
All but one participant report that their companies follow some form of security-by-design process.
Most companies consider security features early in the development process, which is perceived as more effective than addressing them at the end.
Nine interviewees explicitly state that delaying security considerations increases the risk of problems during implementation.

Six participants state that security is often discussed alongside functional features.
Of these, three emphasize the necessity of frequent discussion of security features while realizing functional features, while the other three suggest adding security features into sprint backlogs alongside functional features.
One participant (I23, Developer, ML, Cloud) mentions that they do \textit{``sprints, that are just security focused''} to realize or update security features that have been neglected for a longer period.

Limited by cost, not every company is able to prioritize security features over functional features early on, as reported in ten interviews.
Still, these companies build systems that implement a minimal set of required security features.
\smallskip

\noindent
\textbf{\engineering{threat-model} Threat modeling is widely used in practice}

\result{17}{8}{1}

\sumbox{%
	Many companies employ structured processes based on threat modeling to identify threats to their systems and plan appropriate countermeasures.
	However, the extent and used methodology varies between the individual companies.
}\smallskip

\noindent
\looseness=-1
Threat modeling was identified as the primary security design process, in which developers \textit{``evaluate which threats are there''}\,(I7,\,Security\,Engineer,\,Insurance,\,Banking,\,Retail) determine necessary security features, as described by 17 interviewees.
Threat modeling is also often employed in later development stages when requirements change.
However, threat modeling is often approached in a lightweight way rather than using model-based approaches such as STRIDE\,\cite{shostack2014threat}.
Still, formal model-based threat modeling techniques are mentioned by five participants working in regulated domains such as the automotive or medical industry.
\edit{The identified threats play an important role in selecting security features\,(cf.\,\cref{experience}).}

\looseness=-1
\edit{When participants strongly rely on external service providers such as cloud services to handle their security-critical infrastructure, they limit their threat modeling to identify whether the provided security features are sufficient.}
In these cases, \edit{they inherit the threats from these service providers, but} the responsibility for providing security against them is shifted to the providers---outsourcing risk management is a common practice in standards such as ISO 21434\,\cite{ISO21434}.
\edit{Still, additional threats and thus additional security features may be considered.}
Participants working with customers who do not take security seriously are less likely to apply threat modeling.
\smallskip

\noindent
\textbf{\engineering{baseline} A baseline of standard security features is often implemented but rarely extended and customized}

\result{15}{0}{11}

\sumbox{%
	\looseness=-1
	Companies implement a baseline set of security features required by almost every system, which are only re-engineered as requirements change or vulnerabilities are detected.
}\smallskip

\noindent
Interviewees commonly report that \textit{``there's a set of minimum security features the developers have to implement''} (I21, Manager, Cloud) which is usually derived from company guidelines or regulations.
This baseline typically includes authentication, authorization, validation, logging, and occasionally cryptography.
Once implemented, they are later re-engineered to meet changing requirements, which is often challenging as \textit{``it's always a struggle to search for a library that fulfills the new given requirements''} (I9, Developer, Insurance).

\smallskip

\noindent
\textbf{\engineering{devs} Security features are implemented by ordinary developers as part of their everyday duties}

\result{23}{0}{3}

\sumbox{%
	Developers implement security features in the parts of the system they are responsible for and consult specialized developers for advice rather than having them implement these features.
	Security-specific task assignments are the~exception.
}\smallskip

\noindent
\looseness=-1
All but three interviewees report that ordinary developers without a security background are responsible for implementing security features as a normal part of their tasks.
Developers are expected to identify and implement relevant security features according to the system's security design.
However, some software architects and project managers report they assign security features not to every developer and \textit{``always have an eye on what they are doing''} (I22, Architect, Insurance). 

Most companies train developers to become security experts in certain areas\,(cf.\,\cref{specialization}), allowing them to handle specific security features without needing a dedicated security engineer.
Security engineers with deep security expertise, who aid developers in implementing security features, are only mentioned in four interviews.
Developers proactively seek advice on implementing security features.
Additionally, they are usually tasked with reviewing code written by developers.
\smallskip

\noindent
\textbf{\engineering{externals} Companies rely on secure practices of frameworks and service providers}

\result{20}{0}{6}

\sumbox{%
	Frameworks such as AWS, Azure, or Spring, with best practices and defaults, are widely used and perceived by many to be sufficient for implementing secure systems.
}\smallskip

\noindent
\looseness=-1
Three fourths of the participants explicitly report building their systems based on popular frameworks such as Spring or those provided by cloud providers, e.g., AWS or Azure.
These frameworks often provide secure defaults and best practices, assisting developers in securing software systems.
Following these guidelines helps companies, particularly those with a lower security focus, implement security-by-design principles with low effort.
However, even though participants trust them, one interviewee states that \textit{``you never know what's going on there''} but thinks that they adhere to best practices -- \textit{``at least that's what I hope''} (I2, Manager, Quantum).
\smallskip

\noindent
\textbf{\engineering{test} Security features are often tested, but the majority of testing techniques is not security-specific}

\result{25}{1}{0}

\sumbox{%
	Security features are tested like any other feature, with specific vulnerabilities testing practices rarely applied.
	Since test coverage is a widely used metric, they are often well covered.
	Penetration tests are adopted to a notable degree.
}\smallskip

\noindent
\looseness=-1
Thirteen participants indicate that security features are tested similarly to functional feature through unit, regression and integration tests.
Libraries are usually trusted, but larger companies occasionally test the security of libraries before using them.

Less than half of the interviews mention security-specific testing techniques.
Pentesting is the most common technique, mentioned in nine interviews, usually reported to be cumbersome and ineffective.
Five respondents explain that code reviews or audits by security experts verify the correct implementation and usage of security features.

When explicitly testing security features, three participants report that developers write test cases to verify that unintended behavior does not occur.
For example, access control features are tested to ensure that unauthorized access is properly denied.
\edit{In this case, developers require a \textit{``mindset of 'I want to break things'\,''} (I10, Architect, Medical, Automotive, Robotics).}
However, tests for abuse scenarios are rarely mentioned.
\smallskip

\noindent
\textbf{\engineering{maintenance} Security feature maintenance is \edit{often limited to small fixes and updating dependencies}}

\result{17}{2}{7}

\sumbox{%
	Developers perceive maintaining security features as easy, involving quick fixes or updating libraries.
	Neglecting maintenance leads to accumulating problems and increased effort.
}\smallskip

\noindent
\looseness=-1
Maintenance mainly involves updating security libraries, often supported by automated dependency checks.
Five interviewees mention that \edit{\textit{``it is usually enough to just update the libraries''}\,(I8, Developer, Logistics)}.
Two interviewees rely on external service providers for implementing and updating security features, eliminating the need for maintenance.
Therefore, developers perceive maintaining security features as easy, since tasks such as identifying the code location require low effort.

Due to a lack of security metrics, two interviewees noted that maintenance needs often become apparent only after a vulnerability is discovered or exploited.
Four interviewees report that maintenance effort is high when neglected for long periods, as vulnerabilities accumulate and familiarity fades.
\smallskip

\noindent
\textbf{\engineering{communication} Security features are not communicated systematically and with varying degrees of granularity}

\result{21}{2}{3}

\sumbox{%
	The development phase and stakeholder experience determine how security features are communicated.
	Planning requires a high-level view, implementation low-level technical details.
}\smallskip

\noindent
\looseness=-1
\looseness=-1
Eleven interviewees noted that security features are typically considered at a high level during planning, while twelve stated they must consider details during implementation.
Five interviewees explain developers mainly discuss security concepts, rarely discussing low-level details.
Two developers claim that the lack of implementation details is manageable because \textit{``most of the time you can really work on a high-level abstraction, thanks to the frameworks''}\,(I11, Developer, Systems Management).

\looseness=-1
Selecting appropriate communication granularity strongly depends on the stakeholders involved in a discussion as their knowledge about security features differs.
Five interviewees state that communication to managers or customers is mostly kept on a high level as they lack technical understanding.
\smallskip

\quotebox{Engineering Process of Security Features (RQ2)}{Security features are considered in all phases of software development, and communicated at different levels of granularity.
Even though companies do not employ specific security design processes to model or test security features, most companies employ threat modeling to identify threats.
Initially, security features are realized based on what is commonly done and continuously adapted during development until meeting the requirements.
Selecting service providers supports companies in minimizing security considerations with secure defaults and built-in security mechanisms.
Finally, maintenance tasks for security features are perceived to require little effort.}{5cm}
\section{Characteristics of Security Features (RQ\textsubscript{3})}
\label{sec:characteristics}
\noindent
We now focus on the code-level characteristics of security features, including size (lines of code in relation to other code), tangling (the extent to which a feature intersects/interacts with other features) and scattering (the extent to which a feature is distributed over the codebase) of security features.
\smallskip

\noindent
\textbf{\characteristic{libs} Security features are implemented using frameworks and libraries}

\result{25}{0}{1}

\sumbox{%
	Developers rely on external and internal security frameworks or libraries to avoid implementing vulnerable security features.
	Still, developers write custom code for security features when no suitable library exists or to meet needs.
}\smallskip

\noindent
Participants reported adhering to best practices by using security frameworks and libraries to implement security features. 
Primary motivations are to avoid errors, save costs, and leverage widely used and tested third-party implementations.
For instance, cryptographic algorithms are considered difficult to understand and thus hard to implement. Still, \textit{``you can find a library for everything''}\,(I26, Developer, Manufacturing).

\looseness=-1
However, developers from companies focusing on security solutions occasionally need to write their own code, typically when cutting-edge security features lack libraries: \textit{``Some functions, like for example, if you have some elliptic curves that are not supported, then we implement them on our own, because they do not come with a library yet''} (I4, Manager, Automotive).

Despite existing logging libraries, logging often involves custom implementation\edit{, since developers need to consider what events to log.
The implementation effort is then not related to creating logging functionalities, but to identifying the correct places in the system to add log statements, which leads to more usage than developers perceive for other security features.}
Additionally, many cases of input validation are perceived as specific to the concrete application, requiring a custom implementation tailored to the application-specific needs.
\smallskip

\noindent
\textbf{\characteristic{code} Compared to functional features, the code for security features is minimal}

\result{14}{4}{8}

\sumbox{%
	Since developers rely on libraries, code for security features is limited to their integration using glue code \edit{(code that allows features to interoperate)} and data transformations.
}

\noindent
Our interviews indicate that the size of features in terms of lines of code and thus the effort required to implement them varies, but is kept to a minimum.
Eleven interviewees mention building wrappers around security features to integrate them into the system.
Seven of these claim that this practice facilitates the usage of security libraries by abstracting low-level details.
Still, three interviewees explain that writing wrappers for unfamiliar security features requires additional effort.

In summary, when developing security features, developers primarily write code to pass data to an API of a security library in the correct format or to convert returned data.
For example, two participants mentioned writing wrappers to receive authentication tokens.
In two other interviews, developers explained that creating API wrappers also simplifies data encryption: \textit{``If I have to implement the API wrapper on my own, then I also don't implement the encryption part''} (I8, Developer, Logistics).
To this end, understanding how to integrate security libraries incurs most effort when writing security features.

\smallskip
\noindent
\textbf{\characteristic{modular} Functionality of security features is perceived to be well encapsulated, but usages scattering over the codbase}

\result{20}{3}{3}

\sumbox{%
While the core functionality of security features is implemented in a modular way and can be well-encapsulated, their use is scattered throughout the codebase.
}

\noindent
\looseness=-1
Interviewees report that security features can be well encapsulated and reused throughout the software system.
One interviewee emphasizes: \textit{``you should definitely encapsulate your security code, especially if it's cryptographic code, [\ldots] you don't want to write the same call all over your code, because then if you change something at one place, [\ldots] and then forget about all the others, then your code will have a vulnerability again.''} (I10, Architect, Medical, Automotive, Robotics).
Four participants state that this approach works well for authentication and authorization, while two confirm this practice for cryptography.
\smallskip

\noindent
\textbf{\characteristic{separation} Security features are perceived as being tangled only to a low degree}

\result{16}{4}{6}

\sumbox{%
	Security feature code is perceived as being independent and well separated from other features.
	Developers feel working on security features does not impact other features.
}

\noindent\looseness=-1
Interviewees generally stated that functional features only interact with security features at specific, well-separated locations.
Five participants report that even though functional features interact with security features, changing the security features does not have a direct impact on behavior.
Still, three participants state that they \textit{``try to keep it to a bare minimum, but sometimes you cannot avoid it completely''} (I4, Manager, Automotive).
The only scenario mentioned by interviewees where security feature code is directly woven into the implementation of the functional feature is logging.
One participant perceived a stronger tangling in mainly in the use of cryptographic features such as certificates in the authentication process.
\smallskip

\noindent
\quotebox{\edit{Code-level} characteristics of Security Features (RQ3)}{%
	Developers primarily use libraries for security features, mostly writing wrapper code for integration.
	The functionality of most security features is perceived to be locally encapsulated and well separated from other features' code.
	\edit{Still, the use of these security features is scattered throughout the code base. Counterintuitive to this scattering, security features are perceived as not being tangled with each other.}
}{5cm}
\section{Challenges of Engineering Security Features (RQ\textsubscript{4})}
\label{sec:challenges}
\noindent
Finally, we focus on challenges the participants face in engineering security features and how they overcome them.

\smallskip
\noindent
\textbf{\challenge{knowledge} Non-experts lack foundational knowledge of security}

\result{18}{0}{8}

\sumbox{%
	Insufficient security knowledge among customers and developers hinders security feature engineering, necessitating that companies build a foundation in security knowledge.
}\smallskip

\noindent
To effectively realize security features, \textit{``one requires a basis of knowledge of the security domain''} (I1, Architect, Logistics), which stakeholders usually lack.
Seven interviewees criticize customers' lack of knowledge and concern for security features, with six noting that they only consider them after issues occur.

Four interviewees highlight challenges developers face when implementing security features for the first time, often unsure if they have done so correctly.
One developer reports that library documentation rarely shows examples of misuse or common mistakes, which is \textit{``information I need so that I can write code in a secure way''} (I10, Architect, Medical, Auto\-motive, Robotics).
Consequently, developers are often unsure how security features could be bypassed, because they do not know \textit{``what the attackers do''} (I19, Architect, Banking).

To mitigate this challenge, four participants reported that companies organize workshops and team discussions to improve the general understanding of security features and where people acting in different roles share their knowledge.
However, we observed that many participants were unfamiliar with concepts such as the CIA Triad, preventing them from describing protection goals, therefore highlighting a knowledge gap.

\smallskip

\noindent
\textbf{\challenge{effort} Developers often underestimate the effort of engineering security features}

\result{11}{0}{15}

\sumbox{%
	Lack of knowledge makes it harder for developers to accurately estimate the effort for engineering security features, often leading to complications during implementation.
}\smallskip

\noindent
\looseness=-1
Developers lack an understanding of security features and their underlying concepts.
Before implementing security features, they must first research about them to obtain a basic understanding.
Therefore, two interviewees report it is not possible for them to reliably estimate the effort of realizing security features.

Even after researching security features, developers lack practical experience and conceptual knowledge that is required to effectively implement advanced security features.
Therefore, they often \textit{``initially underestimate the complexity because you only learn about a certain feature when you actually started working on it''} (I21, Manager, Cloud).
In total, six participants report to frequently experience such underestimation, and it often takes longer than expected to realize sophisticated features.
One participant mentions an impact of the development phase, claiming that DevOps security can be estimated well, while estimating implementation effort is challenging.
\smallskip

\noindent
\textbf{\challenge{cost} Developers are unable to implement all security features due to cost constraints}

\result{15}{1}{10}

\sumbox{%
	Developers struggle to implement all security features at an ``optimal'' level due to cost and effort. They struggle to prioritize security features over others to achieve full security.
}\smallskip

\noindent
Seven interviews reveal that developers are not able to implement all security features within a project.
They mention it is impractical to aim for full coverage as \edit{\textit{``you will never be 100\% secure''} (I24, Developer, Manufacturing)}. 
Instead, developers \textit{``try to have a security level that is not so extreme,''} and \textit{``not too expensive to actually implement''} (I4, Manager, Automotive).
One explains, some security features are difficult to prioritize over other features, since they vary in effort.

\looseness=-1
In addition to implementation effort, participants identify learning details as an overhead cost.
This challenge was raised by two participants who explain that developers \edit{need to learn about security features before they are able} to properly implement them, since they are \textit{``not experts in security, but come from another domain''} (I2, Manager, Quantum).

\looseness=-1
Most participants see a barrier to prioritizing security features over functional features, \textit{``because you have to make the whole thing} [the project] \textit{a reality before you find focus on things like that''} (I2, Manager, Quantum).
Three participants state that time constraints prevent them from prioritizing security over functionality.
Customers are an additional impediment, primarily concerned with functionality, which rarely includes security.

\smallskip

\noindent
\textbf{\challenge{estimation} Developers struggle with estimating the side effects of changes to security features}

\result{9}{0}{17}

\sumbox{%
    As security features are used in many parts of the system, estimating how local changes of security features impact the overall system is hard, \edit{since changes to security features are assumed to not impact the functional behavior of the system}.
}\smallskip

\noindent\looseness=-1
One-third of participants report that estimating the side effects of changes is a major challenge when changing security features.
When implementing a new feature, \textit{``it's really hard to see the entire picture [\ldots] and what effects that might have''}\,(I6,\,Developer,\,Real\,Estate).
One of the participants reports that they needed to change access permissions for a new feature, but could not predict how this would affect other existing functionality.
Particularly when functionality which other parties rely on is involved, communication barriers impede the analysis of change impact of security features.
In the worst case, this can lead to outages, e.g., if changes in the authentication services were not properly addressed by consuming parties.
\smallskip

\quotebox{Challenges of Engineering Security Features (RQ4)}{Stakeholders, especially developers, lack knowledge, to estimate the effort required to develop security features.
	Additionally, developers feel overwhelmed learning details about security.
	They struggle to prioritize security features over others and fully implement them because of cost and time constraints.
	Finally, it is challenging to estimate the impact of changes of security features, since they propagate through the system.
}{1.8cm}

\section{Discussion and Common Assumptions}
\label{sec:deductive}
\noindent
We now discuss the findings of our study.

\subsection{Assumptions in relation to practice}
\noindent\looseness=-1
The goal of our study is validating assumptions made in security research.
\Cref{tab:mapping} relates each theme to the assumptions.
\smallskip

\renewcommand{\tabularxcolumn}[1]{m{#1}}
\begin{table*}
	\centering
	\setlength{\tabcolsep}{2.82pt}
	\smaller
	\caption{\edit{Mapping between themes and assumptions {\smaller(\checkmark\hspace{1pt} supports the assumption,
			X\hspace{1.5pt} opposes the assumption)}}}
	\label{tab:mapping}
	\begin{threeparttable}
	\begin{tabularx}{\textwidth}{
			X|
		c>{\columncolor[gray]{0.925}}cc>{\columncolor[gray]{0.925}}cc|>{\columncolor[gray]{0.925}}cc>{\columncolor[gray]{0.925}}cc>{\columncolor[gray]{0.925}}cc>{\columncolor[gray]{0.925}}cc>{\columncolor[gray]{0.925}}c|c>{\columncolor[gray]{0.925}}cc>{\columncolor[gray]{0.925}}c|c>{\columncolor[gray]{0.925}}cc>{\columncolor[gray]{0.925}}c|c>{\columncolor[gray]{0.925}}cc}

	& {\rotatebox{90}{\textbf{\cref{importance} (importance)}}}
	& {\rotatebox{90}{\textbf{\cref{feature} (features)}}}
	& {\rotatebox{90}{\textbf{\cref{experience} (experience-based)}}}
	& {\rotatebox{90}{\textbf{\cref{specialization} (specialization)}}}
	& {\rotatebox{90}{\textbf{\cref{prioritize} (prioritization)}}}
	& {\rotatebox{90}{\textbf{\cref{requirements} (requirements)}}}
	& {\rotatebox{90}{\textbf{\cref{by-design} (by-design)}}}
	& {\rotatebox{90}{\textbf{\cref{threat-model} (threat-model)}}}
	& {\rotatebox{90}{\textbf{\cref{baseline} (baseline)}}}
	& {\rotatebox{90}{\textbf{\cref{devs} (developers)}}}
	& {\rotatebox{90}{\textbf{\cref{externals} (externals)}}}
	& {\rotatebox{90}{\textbf{\cref{test} (testing)}}}
	& {\rotatebox{90}{\textbf{\cref{maintenance} (maintenance)}}}
	& {\rotatebox{90}{\textbf{\cref{communication} (communication)}}}
	& {\rotatebox{90}{\textbf{\cref{libs} (libraries)}}}
	& {\rotatebox{90}{\textbf{\cref{code} (code)}}}
	& {\rotatebox{90}{\textbf{\cref{modular} (modularity)}}}
	& {\rotatebox{90}{\textbf{\cref{separation} (separation)}}}
	& {\rotatebox{90}{\textbf{\cref{knowledge} (knowledge)}}}
	& {\rotatebox{90}{\textbf{\cref{effort} (effort)}}}
	& {\rotatebox{90}{\textbf{\cref{cost} (cost)}}}
	& {\rotatebox{90}{\textbf{\cref{estimation} (side-effects)}}}
	& \multicolumn{3}{c}{
		\begin{tabular}[b]{ccc}
			\multicolumn{3}{c}{\textbf{participant}}\\
			\multicolumn{3}{c}{\textbf{perception}\textsuperscript{1}}\\
			\multicolumn{3}{c}{\textbf{}}\\
			\rotatebox{90}{\textbf{supporting}}
			& \rotatebox{90}{\textbf{opposing}}
			& \rotatebox{90}{\textbf{neither}} \\
		\end{tabular}
	}\\

	\hline
	\textbf{Assumption 1:} Ordinary developers lack knowledge to securely engineer security features (\textsc{Partially Accurate})&  &  & X & \checkmark &  &  &  &  &  & X & \checkmark & \checkmark &  &  & X &  & X &  & \checkmark & \checkmark &  & \checkmark & 4 & 2 & \textbf{20}\\
	\hline
	\textbf{Assumption 2:} (Model-based) security-by-design techniques are not used in practice (\textsc{Not Accurate})& X &  &  &  &  & \checkmark & X & X &  &  & X &  & X & \checkmark &  &  &  &  &  &  &  & & 1 & \textbf{18} & 7\\
	\hline
	\textbf{Assumption 3:} Security libraries are complicated to use, resulting in many vulnerabilities (\textsc{Partially Accurate}) &  &  &  &  &  &  &  &  &  &  &  &  & X &  & X & (X) & (X) & (X) &  & \checkmark &  & & 1 & \textbf{19} & 6\\
	\hline
	\textbf{Assumption 4:} Maximum security is infeasible in practice and security features must be assigned priorities (\textsc{Accurate})& \checkmark & \checkmark &  &  & \checkmark &  &  & \checkmark & \checkmark &  &  &  &  &  &  &  &  &  &  & \checkmark & \checkmark & & \textbf{25} & 0 & 1\\
	\bottomrule
\end{tabularx}
\begin{tablenotes}
	\footnotesize
	\item \textsuperscript{1)} If a 2/3 majority of a participant's agreements or disagreements with the themes related to an assumption support or oppose this assumption, the participant is assumed to support or oppose the assumption.
\end{tablenotes}
\end{threeparttable}
\vspace{-12pt}
\end{table*}

\noindent
\textbf{Assumption 1: Ordinary developers lack knowledge to securely engineer security features (\textsc{Partially Accurate})}

\noindent\looseness=-1
Studies such as Ryan et al.\,\cite{Ryan2023} challenge the underlying assumption that developers’ perceptions of whether security is needed in their work environment are accurate.
Instead, they conclude that developers who are not well-informed about software security may be poor judges of when it is required.
Corresponding observations are also contained in our study, but \edit{reveal that developers often lack conceptual knowledge, yet still have an idea on how to realize security feature practically}.

\looseness=-1
In practice, security features are indeed developed by ordinary developers\,(cf.\,\cref{devs})\edit{, who require experience to effectively select security features\,(cf.\,\cref{experience})}.
However, our interviews revealed a significant lack in security-related knowledge\,(cf.\,\cref{knowledge}) and a tendency to underestimate the effort related to securely engineering security features\,(cf.\,\cref{effort}).
\edit{The lack of knowledge could be one reason for a lack of security-specific test cases\,(cf.\,\cref{test}).}
Particularly, the implementation of detailed security functionalities or their configuration is more complicated, \edit{since a system appears to be working correctly, but may lack important implementation details or usage at specific places of the system\,(cf.\,\cref{estimation})}.
Therefore, specializing in individual security features helps in increasing security and reducing the overhead of learning about all of them\,(cf.\,\cref{specialization}).
The effectiveness of individual experts and the resulting knowledge transfer was also observed in other studies\,\cite{Oyetoyan2016}.
\edit{Relying on service providers that provide secure defaults additionally assists developers in handling the knowledge gap\,(cf.\,\cref{externals}).}
\smallskip

\noindent
\textbf{Assumption 2: (Model-based) security-by-design techniques are not used in practice (\textsc{Not Accurate})}

\noindent

\edit{
	\noindent\looseness=-1
	While researchers from the security-by-design community assume that (model-based) threat modeling techniques are used in practice, other researchers challenge this assumption and paint them as practically irrelevant.
	We lack an understanding of whether (model-based) security-by-design techniques are truly perceived as irrelevant in practice.
}

In the interviews, we find that almost everyone perceives security as important\,(cf.\,\cref{importance}) and incorporates security into their design, although the specifics vary considerably\,(cf.\,\cref{by-design}).
Particularly threat modeling is widely adopted in practice\,(cf.\,\cref{threat-model}).
Companies often consider relying on external service providers or frameworks providing the required functionality through secure defaults\,(cf.\,\cref{externals}), acknowledging the inheritance of potential threats that may need to be monitored\,(cf.\,\cref{maintenance}).
Even when interviewees stated that security is not a priority, we learned that there are ``security sprints'' or security features that have been developed from the start\,(cf.\,\cref{by-design}).
Security-by-design is, therefore, widely adopted in practice, although sometimes not perceived as such, and usually much less systematically than techniques such as UMLsec\,\cite{Juerjens05} or STRIDE\,\cite{shostack2014threat}.
\edit{However, concrete security requirements are usually lacking at project start, leading to challenges, especially in unregulated domains\,(cf.\,\cref{requirements}).
Finally, we found that security features are not systematically communicated\,(cf.\,\cref{communication}) or captured in models\,(cf.\,\cref{threat-model}) throughout the development process.
Our interviews highlight a need for existing techniques to be more practical for effective usage in the development process.}
\smallskip

\noindent
\textbf{Assumption 3: Security libraries are complicated to use, resulting in many vulnerabilities (\textsc{Partially Accurate})}

\noindent
\looseness=-1
Insecure use of cryptographic APIs has been identified as a common reason for insecure code.
Tools such as CogniCrypt\,\cite{KNR+2017} address this by providing automated checks \edit{and code generation for} usages of cryptographic APIs.
It is a common conception that the documentation of libraries is a major factor in secure use, albeit often considered inadequate\,\cite{Zhong2013}.

In general, participants explain that they use libraries wherever possible\,(cf.\,\cref{libs}).
In contrast to common assumptions\,\cite{Zhong2013,Piccioni2013,Myers2016}, the interviewees of our study generally were satisfied with the quality of provided APIs and their documentation, claiming they can be easily integrated with minimal effort\,(cf.\,\cref{code}), well separated from other parts of the system\,(cf.\,\cref{modular}\,and\,\cref{separation}), and easily updated when required\,(cf.\,\cref{maintenance}).
However, studies have shown that even if developers use only libraries, the written code can be insecure due to their incorrect usage\,\cite{Acar2017}.
We learned from the interviews that \edit{incorrect use is often difficult to identify\,(cf.\,\cref{estimation}), leading to a discrepancy between perceived ease and practical challenge.}
Ultimately, developers perceive implementing security features as easy, while it is difficult in reality\,(cf.\,\cref{effort}).
Developers need a general understanding of how a library works internally, but do not want or can go deeply into the algorithmic details.
Consequently, developers may oversimplify implementations, potentially introducing vulnerabilities.
Although, we asked participants in which security features they identify the most bugs, the answers did not reveal a pattern or participants were unable to confidently provide an answer.
\smallskip

\noindent
\textbf{Assumption 4: Maximum security is infeasible in practice and security features must be assigned priorities (\textsc{Accurate})}

\noindent
\looseness=-1
Since complete security is not feasible, assigning priorities to security features is required\,\cite{Mazurek2022}.
The participants confirm \edit{that security is an important aspect, but since functionality is more critical\,(cf.\,\cref{importance})}, only limited resources are available and they cannot realize every security feature\,(cf.\,\cref{cost}).
Other studies find that 20\,\% of the median effort for new projects is due to security\,\cite{Venson2019}.
Prioritization of security features is an essential part of the development process in practice\,(cf.\,\cref{prioritize}), \edit{often involving the consideration of different threats\,(cf.\,\cref{threat-model})}.
Since the task of assigning priorities to security features must be at least partially performed by non-expert developers, often doing many of these things for the first time, they often underestimate the effort involved in realizing them\,(cf.\,\cref{effort}).
Since the selection of security features is based on experience\,(cf.\,\cref{experience}), developers often struggle in selecting security features beyond the common baseline\,(cf.\,\cref{feature}\,and\,\cref{baseline}).
This makes effort estimation difficult, which makes assigning priorities difficult.
Ultimately, this might result in important security features not being implemented because of other priorities.

\subsection{Findings and Observations}
	\noindent
	We provide insights into the engineering of security features:

	\noindent
	\textbf{Finding 1.} Security is a cross-cutting concern that affects many domains, but requires specific security knowledge.
	In line with the intuition of researchers, this knowledge is only available to and recognized by a few developers.
	However, counterintuitively, developers still perceive themselves to be able to implement security features correctly because security libraries seem to be easy to use.
	Still, when comparing with previous work\,\cite{Acar2017}, it is uncertain whether developers truly implement them correctly, since a software system may operate functionally correct even when security features are missing.

	\noindent
	\textbf{Finding 2.} Systematically engineering security features requires specific processes, e.g., threat modeling, that are orthogonal to software development.
	While companies integrate such activities into their development processes, they mainly integrate them without changing the process.
	Counterintuitively, this leads to activities such as threat modeling being used in practice in a lightweight manner tailored towards existing processes.

	\noindent
	\textbf{Finding 3.} While intuition suggests that implementing security features involves huge effort, security feature implementation is perceived to require little code to integrate libraries.
	They are perceived to be well encapsulated, rarely tangled, and easy to maintain.
	Counterintuitively, estimating side effects is perceived especially difficult, since it is unclear what parts are affected by a change.
	Overall, proper planning of needed security features is particularly challenging.
	Therefore, security-by-design methodologies should be automated to enable developers with limited skills to perform security assessment quickly\,\cite{casola2020securityautomation}.
\subsection{Implications}
\noindent
Our study highlights implications on the engineering of security features for both researchers and practitioners.

\looseness=-1
\parhead{Security knowledge.} We notice a gap in foundational security knowledge among developers and stakeholders.
Therefore, they require better tools and resources that help them integrate security features early in the development process.
\textit{Researchers} should focus on creating documentation including both correct usage examples and common misuse cases.
\edit{\textit{Practitioners} need to build deeper security knowledge to be able to notice insecure practices by e.g., adopting the concept of lighthouses.}

\looseness=-1
\parhead{Prioritization.} The difficulty of balancing security and functionality with limited project resources is a significant challenge.
\textit{Researchers} need to explore methods for effectively prioritizing security features, both in terms of functionality and against each other, without compromising essential functionality.
This includes developing strategies that assist in making informed decisions about security investments.
\edit{\textit{Practitioners} implicitly prioritize security features continuously, but should explicitly document decisions to facilitate reasoning and security hardening.}

\looseness=-1
\parhead{Security metrics.} The lack of security metrics hampers the ability to assess clear and actionable insights about the security level of a system.
Effective security metrics could not only assist developers in assigning priorities for security features, but also encourage stakeholders in understanding risks of neglecting security features.
We identify a pressing need for security metrics which \textit{researchers} should further explore.
\edit{\textit{Practitioners} could adopt formal security-by-design techniques which allow a systematic reasoning about a system's security\,\cite{Juerjens05,Shostack2008,Tuma2019}}.

\looseness=-1
\parhead{Modeling techniques.} Threat modeling is a crucial practice among developers, yet often performed in an informal and light-weight way.
The \textit{research community} should therefore investigate more efficient and accessible modeling techniques, that aid developers in integrating them into their workflows.

\section{Threats to Validity}
\label{sec:validity}
\noindent

\parhead{Internal Validity.}
Author bias could impact our results.
Since we used semi-structured interviews, some questions might have been missed.
We employed two authors to conduct each interview, both ensuring all questions were covered.
To reduce participant bias, we mainly used open-ended questions, resorting to closed-ended ones only for clarification.
We gave specific examples only if participants struggled to answer, thus minimizing bias and fostering understanding.
The first author coded the interviews, but two authors thoroughly discussed the code book to validate and counteract any potential biases.

\looseness=-1
\parhead{External Validity.}
The generalizability of our results might be threatened due to the composition of participants\,\cite{Elston2021}.
Company standards influence participants' experiences. We chose companies from diverse domains with varying security importance, finding security relevance in nearly all.
Participants' backgrounds may affect generalizability, \edit{since willingness to participate depends on self-perceived experience\,\cite{Elston2021}}.
Therefore, we recruited individuals with varied experiences and roles to ensure diversity.
\edit{Response} bias could threaten generalizability, when participants are hesitant to admit insecure practices\,\cite{Elston2021}.
We thus contrasted and validated interviewee perceptions with other empirical studies, especially where conflicts with empirical findings arose.
Furthermore, the inability of participants to answer all questions \edit{could limit the accuracy of answers due to response bias\,\cite{Elston2021}}. We still received detailed responses, covering all research questions and reaching saturation.
\edit{Lastly, our study provides insights on the engineering of security features from developer perspective.
Since we did not have access to the participants' projects, we were unable to triangulate them with additional artifacts.
Future research should investigate discrepancies between this perception and actual implementation of security features by analyzing software systems.}

\section{Conclusion}
\label{sec:conclusion}
\noindent
\looseness=-1
\edit{While the research community provides security technologies and usage guidelines with many underlying assumptions based on intuition, little is known about how developers face challenges in practice when developing security features.}
To address this gap, we contribute an exploratory interview study with 26 experts from industry.
We elicited their experience to validate and contextualize four common assumptions.

\looseness=-1
Our study reveals that companies generally follow security-by-design principles, yet specific security requirements are sparse.
Threat modeling is widely used to identify potential threats and devise corresponding security measures.
Security features are routinely included in most software systems by using established frameworks and libraries, adhering to best practices and default settings. 
Developers perceive security features as easy to implement, effectively separating security features from other functionalities and minimizing the amount of code.


\looseness=-1
Nonetheless, developers struggle with engineering security features due to a lack of foundational knowledge about them, often underestimating the effort needed.
To address these challenges, developers often undergo training in specific areas of security.
However, time constraints force them to prioritize functional features over security enhancements, hampering the comprehensive implementation of necessary security measures.

\looseness=-1
\edit{
As our study has shown, many assumptions in security research may not be empirically validated and partially or even entirely inaccurate.
Researchers should aim to validate more assumptions empirically by, e.g., conducting surveys and interviews with developers, or case studies of software projects, to ensure that their theories match practice.
}

\section*{Acknowledgements}\noindent
We thank all 26 interviewees for their participation. This work was partially funded by the Deutsche Forschungsgemeinschaft (DFG, German Research Foundation) under Germany's Excellence Strategy - EXC 2092 CASA - 390781972.

\balance{}
\bibliographystyle{IEEEtran}
\bibliography{IEEEabrv,references}

\begin{thebibliography}{10}
\providecommand{\url}[1]{#1}
\csname url@samestyle\endcsname
\providecommand{\newblock}{\relax}
\providecommand{\bibinfo}[2]{#2}
\providecommand{\BIBentrySTDinterwordspacing}{\spaceskip=0pt\relax}
\providecommand{\BIBentryALTinterwordstretchfactor}{4}
\providecommand{\BIBentryALTinterwordspacing}{\spaceskip=\fontdimen2\font plus
\BIBentryALTinterwordstretchfactor\fontdimen3\font minus \fontdimen4\font\relax}
\providecommand{\BIBforeignlanguage}[2]{{%
\expandafter\ifx\csname l@#1\endcsname\relax
\typeout{** WARNING: IEEEtran.bst: No hyphenation pattern has been}%
\typeout{** loaded for the language `#1'. Using the pattern for}%
\typeout{** the default language instead.}%
\else
\language=\csname l@#1\endcsname
\fi
#2}}
\providecommand{\BIBdecl}{\relax}
\BIBdecl

\bibitem{McGraw2004}
G.~McGraw, ``{Software Security},'' \emph{IEEE Security \& Privacy}, vol.~2, no.~2, pp. 80--83, 2004.

\bibitem{Cawthra2020}
J.~Cawthra, M.~Ekstrom, L.~Lusty, J.~Sexton, and J.~Sweetnam, ``{Data Integrity: Detecting and Responding to Ransomware and Other Destructive Events},'' National Institute of Standards and Technology, Tech. Rep. NIST SP 1800-26, 2020.

\bibitem{Parker2015}
D.~B. Parker, ``{Toward a New Framework for Information Security},'' in \emph{{Computer Security Handbook}}, 1st~ed., S.~Bosworth, M.~E. Kabay, and E.~Whyne, Eds.\hskip 1em plus 0.5em minus 0.4em\relax Wiley, 2015, vol.~1, ch.~3, pp. 3.1--3.23.

\bibitem{Sindre2005}
G.~Sindre and A.~L. Opdahl, ``Eliciting security requirements with misuse cases,'' \emph{Requirements Engineering}, vol.~10, no.~1, pp. 34--44, 2005.

\bibitem{Haley2008}
C.~B. Haley, R.~C. Laney, J.~D. Moffett, and B.~Nuseibeh, ``{Security Requirements Engineering: {A} Framework for Representation and Analysis},'' \emph{Transactions on Software Engineering (TSE)}, vol.~34, no.~1, pp. 133--153, 2008.

\bibitem{Mellado2010}
D.~Mellado, C.~Blanco, L.~E. S{\'{a}}nchez, and E.~Fern{\'{a}}ndez{-}Medina, ``A systematic review of security requirements engineering,'' \emph{Computer Standards \& Interfaces}, vol.~32, no.~4, pp. 153--165, 2010.

\bibitem{Juerjens05}
J.~J{\"{u}}rjens, \emph{Secure Systems Development with {UML}}.\hskip 1em plus 0.5em minus 0.4em\relax Springer, 2005.

\bibitem{shostack2014threat}
A.~Shostack, \emph{{Threat Modeling: Designing for Security}}.\hskip 1em plus 0.5em minus 0.4em\relax John Wiley \& Sons, 2014.

\bibitem{SDG2017}
M.~Salnitri, F.~Dalpiaz, and P.~Giorgini, ``Designing secure business processes with {SecBPMN},'' \emph{Software \& Systems Modeling (SoSyM)}, vol.~16, no.~3, pp. 737--757, 2017.

\bibitem{casola2020securityautomation}
V.~Casola, A.~{De Benedictis}, M.~Rak, and U.~Villano, ``A novel security-by-design methodology: {M}odeling and assessing security by {SLAs} with a quantitative approach,'' \emph{Journal of Systems and Software (JSS)}, vol. 163, 2020.

\bibitem{green2016developers}
M.~Green and M.~Smith, ``Developers are not the enemy!: {T}he need for usable security {APIs},'' \emph{IEEE Security {\&} Privacy}, vol.~14, no.~5, pp. 40--46, 2016.

\bibitem{potter2004}
B.~Potter and G.~McGraw, ``Software security testing,'' \emph{IEEE Security \& Privacy}, vol.~2, no.~5, pp. 81--85, 2004.

\bibitem{PTS+2019}
S.~Peldszus, K.~Tuma, D.~Str{\"{u}}ber, J.~J{\"{u}}rjens, and R.~Scandariato, ``{Secure Data-Flow Compliance Checks between Models and Code based on Automated Mappings},'' in \emph{International Conference on Model-driven Engineering Languages and Systems (MODELS)}, 2019, pp. 23--33.

\bibitem{TPS2022}
K.~Tuma, S.~Peldszus, D.~Strüber, R.~Scandariato, and J.~Jürjens, ``{Checking Security Compliance between Models and Code},'' \emph{International Journal on Software and Systems Modeling (SoSyM)}, 2022.

\bibitem{di2023}
D.~Di~Dario, V.~Pontillo, S.~Lambiase, F.~Ferrucci, F.~Palomba \emph{et~al.}, ``Security testing in the wild: {An} interview study,'' in \emph{Euromicro Conference on Software Engineering and Advanced Applications (SEAA)}, 2023, pp. 191--198.

\bibitem{Ryan2023}
I.~Ryan, U.~Roedig, and K.~Stol, ``Unhelpful assumptions in software security research,'' in \emph{Conference on Computer \& Communications Security (CCS)}, 2023, pp. 3460--3474.

\bibitem{Wurster2008}
G.~Wurster and P.~Oorschot, ``{The Developer is the Enemy},'' in \emph{New Security Paradigms Workshop}, 01 2008, pp. 89--97.

\bibitem{KNR+2017}
S.~Kr{\"{u}}ger, S.~Nadi, M.~Reif, K.~Ali, M.~Mezini, E.~Bodden, F.~G{\"{o}}pfert, F.~G{\"{u}}nther, C.~Weinert, D.~Demmler, and R.~Kamath, ``{CogniCrypt}: {S}upporting developers in using cryptography,'' in \emph{International Conference of Automated Software Engineering (ASE)}, 2017, pp. 931--936.

\bibitem{WBBM2021}
\BIBentryALTinterwordspacing
A.-K. Wickert, L.~Baumg\"{a}rtner, F.~Breitfelder, and M.~Mezini, ``Python crypto misuses in the wild,'' in \emph{International Symposium on Empirical Software Engineering and Measurement (ESEM)}, 2021. [Online]. Available: \url{https://doi.org/10.1145/3475716.3484195}
\BIBentrySTDinterwordspacing

\bibitem{egele2013empirical}
M.~Egele, D.~Brumley, Y.~Fratantonio, and C.~Kruegel, ``An empirical study of cryptographic misuse in android applications,'' in \emph{Conference on Computer \& Communications Security (CCS)}, 2013, pp. 73--84.

\bibitem{Acar2017}
Y.~Acar, M.~Backes, S.~Fahl, S.~Garfinkel, D.~Kim, M.~L. Mazurek, and C.~Stransky, ``Comparing the usability of cryptographic {APIs},'' in \emph{Symposium on Security and Privacy (SP)}, 2017, pp. 154--171.

\bibitem{Oyetoyan2016}
T.~D. Oyetoyan, D.~S. Cruzes, and M.~G. Jaatun, ``An empirical study on the relationship between software security skills, usage and training needs in agile settings,'' in \emph{International Conference on Availability, Reliability and Security (ARES)}, 2016, pp. 548--555.

\bibitem{nadi2016jumping}
S.~Nadi, S.~Kr{\"u}ger, M.~Mezini, and E.~Bodden, ``Jumping through hoops: why do {J}ava developers struggle with cryptography {APIs?}'' in \emph{International Conference on Software Engineering (ICSE)}, 2016, pp. 935--946.

\bibitem{Venson2019}
E.~Venson, R.~Alfayez, M.~M.~F. Gomes, R.~M. da~C.~Figueiredo, and B.~W. Boehm, ``The impact of software security practices on development effort: An initial survey,'' in \emph{International Symposium on Empirical Software Engineering and Measurement (ESEM)}, 2019, pp. 1--12.

\bibitem{ReplicationPackage}
``Replication package,'' \url{https://doi.org/10.5281/zenodo.14237228}, 2025.

\bibitem{berger2015feature}
T.~Berger, D.~Lettner, J.~Rubin, P.~Gr{\"u}nbacher, A.~Silva, M.~Becker, M.~Chechik, and K.~Czarnecki, ``What is a feature? a qualitative study of features in industrial software product lines,'' in \emph{International Systems and Software Product Line Conference (SPLC)}, 2015.

\bibitem{bosch2000}
J.~Bosch, \emph{Design \& Use of Software Architectures—{A}dopting and Evolving a Product Line Approach}.\hskip 1em plus 0.5em minus 0.4em\relax Addison-Wesley, 01 2000.

\bibitem{Batory:2004bw}
D.~Batory, J.~N. Sarvela, and A.~Rauschmayer, ``{Scaling Step-Wise Refinement},'' \emph{Transactions on Software Engineering (TSE)}, vol.~30, no.~6, pp. 355--371, 2004.

\bibitem{chen2005}
K.~Chen, W.~Zhang, H.~Zhao, and H.~Mei, ``An approach to constructing feature models based on requirements clustering,'' in \emph{International Conference on Requirements Engineering (RE)}, USA, 2005, p. 31–40.

\bibitem{kang1990}
K.~Kang, S.~Cohen, J.~Hess, W.~Novak, and A.~Peterson, ``Feature-oriented domain analysis {(FODA)} feasibility study,'' Software Engineering Institute, Carnegie Mellon University, Pittsburgh, PA, Tech. Rep. CMU/SEI-90-TR-021, 1990.

\bibitem{riebisch2003}
M.~Riebisch, ``Towards a more precise definition of feature models,'' \emph{Modelling Variability for Object-Oriented Product Lines}, 2003.

\bibitem{hermann2025}
\BIBentryALTinterwordspacing
K.~Hermann, S.~Schneider, C.~Tony, A.~Yardim, S.~Peldszus, T.~Berger, R.~Scandariato, M.~A. Sasse, and A.~Naiakshina, ``A taxonomy of functional security features and how they can be located,'' 2025. [Online]. Available: \url{https://arxiv.org/abs/2501.04454}
\BIBentrySTDinterwordspacing

\bibitem{tsipenyuk2005}
K.~Tsipenyuk, B.~Chess, and G.~McGraw, ``Seven pernicious kingdoms: a taxonomy of software security errors,'' \emph{IEEE Security \& Privacy}, vol.~3, no.~6, pp. 81--84, 2005.

\bibitem{Bass2003}
L.~Bass, P.~Clements, and R.~Kazman, \emph{Software Architecture In Practice}.\hskip 1em plus 0.5em minus 0.4em\relax Addison-Wesley Longman, 01 2003.

\bibitem{Allen2008}
J.~H. Allen, S.~Barnum, R.~J. Ellison, G.~McGraw, and N.~R. Mead, \emph{Software Security Engineering}.\hskip 1em plus 0.5em minus 0.4em\relax Pearson Education, 2008.

\bibitem{schumacher2013}
M.~Schumacher, E.~Fernandez-Buglioni, D.~Hybertson, F.~Buschmann, and P.~Sommerlad, \emph{Security Patterns: Integrating security and systems engineering}.\hskip 1em plus 0.5em minus 0.4em\relax John Wiley \& Sons, 2013.

\bibitem{Peldszus2022}
S.~Peldszus, \emph{{Security Compliance in Model-driven Development of Software Systems in Presence of Long-Term Evolution and Variants}}.\hskip 1em plus 0.5em minus 0.4em\relax Springer, 2022.

\bibitem{ryoo2016}
J.~Ryoo, B.~Malone, P.~A. Laplante, and P.~Anand, ``The use of security tactics in open source software projects,'' \emph{IEEE Transactions on Reliability}, vol.~65, no.~3, pp. 1195--1204, 2016.

\bibitem{passos2015scattering}
L.~Passos, J.~Padilla, T.~Berger, S.~Apel, K.~Czarnecki, and M.~T. Valente, ``Feature scattering in the large: A longitudinal study of linux kernel device drivers,'' in \emph{14th International Conference on Modularity (MODULARITY)}, 2015.

\bibitem{apel2013featureinteractions}
S.~Apel, D.~Batory, C.~K{\"a}stner, and G.~Saake, \emph{Feature Interactions}.\hskip 1em plus 0.5em minus 0.4em\relax Berlin, Heidelberg: Springer Berlin Heidelberg, 2013, pp. 213--241.

\bibitem{Werlinger2009}
R.~Werlinger, K.~Hawkey, and K.~Beznosov, ``An integrated view of human, organizational, and technological challenges of it security management,'' \emph{Information Management \& Computer Security}, vol.~17, no.~1, pp. 4--19, 2009.

\bibitem{Klivan2024}
S.~Klivan, S.~Höltervennhoff, R.~Panskus, K.~Marky, and S.~Fahl, ``Everyone for themselves? a qualitative study about individual security setups of open source software contributors,'' in \emph{2024 IEEE Symposium on Security and Privacy (SP)}, 2024, pp. 1065--1082.

\bibitem{shull2007}
F.~Shull, J.~Singer, and D.~I. Sj\o{}berg, \emph{Guide to Advanced Empirical Software Engineering}.\hskip 1em plus 0.5em minus 0.4em\relax Springer-Verlag, 2007.

\bibitem{denzin2011}
N.~K. Denzin and Y.~S. Lincoln, \emph{The {Sage} handbook of qualitative research}.\hskip 1em plus 0.5em minus 0.4em\relax Sage, 2011.

\bibitem{Weller2018}
S.~C. Weller, B.~Vickers, H.~R. Bernard, A.~M. Blackburn, S.~Borgatti, C.~C. Gravlee, and J.~C. Johnson, ``{Open-Ended Interview Questions and Saturation},'' \emph{PLOS ONE}, vol.~13, no.~6, 2018.

\bibitem{Hennink2017}
M.~M. Hennink, B.~N. Kaiser, and V.~C. Marconi, ``{Code Saturation Versus Meaning Saturation: How Many Interviews Are Enough?}'' \emph{Qualitative Health Research}, vol.~27, no.~4, pp. 591--608, 2017.

\bibitem{Whisper}
``Whisper,'' \url{https://github.com/openai/whisper}.

\bibitem{Boyatzis1998}
R.~E. Boyatzis, \emph{Transforming Qualitative Information}.\hskip 1em plus 0.5em minus 0.4em\relax sage, 1998.

\bibitem{MAXQDA}
``{MAXQDA Website},'' \url{https://www.maxqda.com/}, 2024.

\bibitem{Ahmadian2017}
A.~S. Ahmadian, S.~Peldszus, Q.~Ramadan, and J.~J{\"{u}}rjens, ``{Model-based privacy and security analysis with CARiSMA},'' in \emph{Joint Meeting on Foundations of Software Engineering (ESEC/FSE)}, E.~Bodden, W.~Sch{\"{a}}fer, A.~van Deursen, and A.~Zisman, Eds.\hskip 1em plus 0.5em minus 0.4em\relax {ACM}, 2017, pp. 989--993.

\bibitem{Reiche2024}
F.~Reiche, T.~Weber, S.~Becker, S.~Weber, R.~Heinrich, and E.~Burger, ``{Consistency Management for Security Annotations for Continuous Verification},'' in \emph{International Conference on Model Driven Engineering Languages and Systems (MODELS)}, M.~Wimmer, A.~Egyed, B.~Combemale, and M.~Chechik, Eds.\hskip 1em plus 0.5em minus 0.4em\relax {ACM}, 2024, pp. 1096--1105.

\bibitem{Krausz2024}
M.~Krausz, S.~Peldszus, F.~Regazzoni, T.~Berger, and T.~G{\"{u}}neysu, ``{120 Domain-Specific Languages for Security},'' \emph{CoRR}, vol. abs/2408.06219, 2024.

\bibitem{tuma2018threat}
K.~Tuma, G.~Calikli, and R.~Scandariato, ``{Threat Analysis of Software Systems: A Systematic Literature Review},'' \emph{Journal of Systems and Software (JSS)}, vol. 144, pp. 275--294, 2018.

\bibitem{TSB2019}
K.~Tuma, R.~Scandariato, and M.~Balliu, ``Flaws in flows: {U}nveiling design flaws via information flow analysis,'' in \emph{International Conference on Software Architecture (ICSA)}, 2019, pp. 191--200.

\bibitem{Apvrille2005}
A.~Apvrille and M.~Pourzandi, ``Secure software development by example,'' \emph{{IEEE} Security and Privacy}, vol.~3, no.~4, pp. 10--17, 2005.

\bibitem{Best2007}
B.~Best, J.~J{\"{u}}rjens, and B.~Nuseibeh, ``Model-based security engineering of distributed information systems using {UMLsec},'' in \emph{International Conference on Software Engineering (ICSE)}, 2007, pp. 581--590.

\bibitem{Juerjens2008}
J.~J{\"{u}}rjens, J.~Schreck, and P.~Bartmann, ``Model-based security analysis for mobile communications,'' in \emph{International Conference on Software Engineering (ICSE)}, 2008, pp. 683--692.

\bibitem{Basin2011}
D.~A. Basin, M.~Clavel, and M.~Egea, ``{A Decade of Model-Driven Security},'' in \emph{Symposium on Access Control Models and Technologies (SACMAT)}, 2011, pp. 1--10.

\bibitem{Seifermann2019}
S.~Seifermann, R.~Heinrich, and R.~H. Reussner, ``{Data-Driven Software Architecture for Analyzing Confidentiality},'' in \emph{International Conference on Software Architecture (ICSA)}, 2019, pp. 1--10.

\bibitem{Peldszus2021}
S.~Peldszus, J.~Bürger, T.~Kehrer, and J.~Jürjens, ``Ontology-driven evolution of software security,'' \emph{Data \& Knowledge Engineering (DKE)}, vol. 134, 2021.

\bibitem{Peldszus2024}
S.~Peldszus, J.~B{\"{u}}rger, and J.~J{\"{u}}rjens, ``{UMLsecRT}: {R}eactive security monitoring of java applications with round-trip engineering,'' \emph{Transactions on Software Engineering (TSE)}, vol.~50, no.~1, pp. 16--47, 2024.

\bibitem{Gorschek2014}
T.~Gorschek, E.~Tempero, and L.~Angelis, ``On the use of software design models in software development practice: An empirical investigation,'' \emph{Journal of Systems and Software (JSS)}, vol.~95, pp. 176--193, 2014.

\bibitem{Mazurek2022}
M.~L. Mazurek, ``We are the experts, and we are the problem: {The} security advice fiasco,'' in \emph{Conference on Computer \& Communications Security (CCS)}, 2022.

\bibitem{ISO21434}
I.~S. 32, ``Road vehicles -- cybersecurity engineering,'' International Organization for Standardization (ISO), International Standard ISO/SAE 21434, 2021.

\bibitem{Zhong2013}
H.~Zhong and Z.~Su, ``Detecting {API} documentation errors,'' in \emph{International Conference on Object Oriented Programming Systems Languages {\&} Applications (OOPSLA)}, 2013, pp. 803--816.

\bibitem{Piccioni2013}
M.~Piccioni, C.~A. Furia, and B.~Meyer, ``An empirical study of {API} usability,'' in \emph{International Symposium on Empirical Software Engineering and Measurement (ESEM)}, 2013, pp. 5--14.

\bibitem{Myers2016}
B.~A. Myers and J.~Stylos, ``Improving {API} usability,'' \emph{Communications of the ACM}, vol.~59, no.~6, p. 62–69, 2016.

\bibitem{Shostack2008}
A.~Shostack, ``Experiences threat modeling at microsoft,'' in \emph{MODSEC}, 2008.

\bibitem{Tuma2019}
K.~Tuma, R.~Scandariato, and M.~Balliu, ``Flaws in flows: Unveiling design flaws via information flow analysis,'' in \emph{2019 IEEE International Conference on Software Architecture (ICSA)}, 2019, pp. 191--200.

\bibitem{Elston2021}
D.~M. Elston, ``Participation bias, self-selection bias, and response bias,'' \emph{Journal of the American Academy of Dermatology}, 2021.

\end{thebibliography}

\end{document}